\documentclass[acmsmall, nonacm]{acmart}
\AtBeginDocument{%
  \providecommand\BibTeX{{%
    \normalfont B\kern-0.5em{\scshape i\kern-0.25em b}\kern-0.8em\TeX}}}

\begin{document}

\title{Misinformation as Information Pollution}

\author{Ashkan Kazemi}
\email{ashkank@umich.edu}
\orcid{0000-0002-2475-1007}
\author{Rada Mihalcea}
\email{mihalcea@umich.edu}
\orcid{0000-0002-0767-6703}
\affiliation{%
  \institution{University of Michigan}
  \streetaddress{2260 Hayward St.}
  \city{Ann Arbor}
  \state{Michigan}
  \country{USA}
  \postcode{48109-2121}
}

\renewcommand{\shortauthors}{Kazemi and Mihalcea}

\begin{abstract}
  Social media feed algorithms are designed to optimize online social engagements for the purpose of maximizing advertising profits, and therefore have an incentive to promote controversial posts including misinformation. 
  By thinking about misinformation as information pollution, we can draw parallels with environmental policy for countering pollution such as carbon taxes. Similar to pollution, a Pigouvian tax on misinformation provides economic incentives for social media companies to control the spread of misinformation more effectively to avoid or reduce their misinformation tax, while preserving some degree of freedom in platforms’ response.
  In this paper, we highlight a bird’s eye view of a Pigouvian misinformation tax and discuss the key questions and next steps for implementing such a taxing scheme.
\end{abstract}





\maketitle

\section{Introduction}
 Pollution is the introduction of harmful materials into the environment, either naturally such as volcanic ash or introduced by human activity such as runoff produced by factories\footnote{\url{https://education.nationalgeographic.org/resource/pollution/}}. 
 Pollutants damage our air quality, water and the environment at large. We apply the famous “duck test\footnote{\url{https://en.wikipedia.org/wiki/Duck_test}}” to pollution and misinformation: if misinformation looks like pollution and acts like pollution, then it is \textit{information} pollution. 
 Similar to toxic water pollutants that poison city drinking waters, the spread of information pollution into our online social lives causes damages to individuals and societies that are sometimes beyond repair. Online misinformation has ignited catastrophic social distress in recent years such as a genocide in Myanmar \cite{myanmar} and worsening global public health during the COVID-19 pandemic \cite{caceres2022impact}.
 As with environmental regulation policies around pollutants, when a company causes leaks into the environment, whether this is caused by an individual’s negligence or structural and cultural faults of the company, the company is responsible for the environmental damage.

 There are three main structural incentives in the social media economy that promote misinformation: \textbf{(i)} maximizing user engagement by all means necessary promotes controversial content such as misinformation as such posts are likely to yield higher user engagement \cite{vosoughi2018spread}, \textbf{(ii)} Algorithms can identify the right influencers for maximizing advertisement reach \cite{ahmadinejad2014effectively}, or marketing firms use AB-tests to tailor ads to their target audiences \cite{wilson2019cambridge}; in such settings adversarial actors can make use of such algorithms to spread misinformation more effectively, \textbf{(iii)} algorithms are not penalized for suggesting harmful content but receive rewards (i.e. ad revenue). The lack of penalty affords feed algorithms a larger playground for exploration that might include recommendation for extremist Facebook groups or Youtube channels.

 Adopting an environmentally inspired tax on misinformation, directed at the social media platforms, can curb the spread of misinformation.
 A Pigouvian tax \cite{pigou2017economics} suggests taxing negative externalities by setting a tax rate equal to the social marginal damages resulting from an additional unit of the externality. Different variations of this tax have been suggested to counter the negative externalities of carbon emissions \cite{metcalf2009design, metcalf2009designing}, tobacco \cite{cnossen2006tobacco, chaloupka2012tobacco}, alcohol \cite{cnossen2007alcohol, bouchery2011economic, wagenaar2010effects} and traffic congestion \cite{lindsney2001traffic, parry2002comparing}.
 Few proposals to implement a social media or fake news tax have been proposed: the Nobel-winning economist Paul Romer proposed that a tax on targeted advertising revenue would incentivize ``big tech’’ to shift their dangerous business model away from misinformation and hate speech \cite{bigtechtax} . In response to arguments against this tax, Romer noted that even if tech companies move away from this economy and make the tax obsolete, this would still bring in real change as the companies have found alternative revenue streams and therefore are not as prone to spread and promotion of toxic content. A similar proposal by Van Alstyne \cite{bufakenewstax} suggests imposing monetary penalties to fake news posts and the individuals who created them, while also keeping the integrity of freedom of speech. Additionally, experimental evidence from \cite{rathje2023accuracy} suggests economic incentives and more broadly motivation based. An incentive as little as a dollar for forming accurate judgements improved accuracy in judgments of political news headlines in the study’s participants. At macro scale, research suggests that belief in conspiracies is strongly related to helplessness \cite{doi:10.1126/science.1159845} and inequality in relative socioeconomic status \cite{payne2018broken} in societies.

 In this paper we contribute (i) a high-level framework for misinformation tax, (ii) discuss existing evidence and estimates of the financial and public health burdens of misinformation, and (iii) provide additional considerations for what such a tax might look like in practice.

\section{What counts as \textit{misinformation?}}
We define misinformation to be misleading information shared online, regardless of intent. While several other terms have been used to address the phenomenon such as fake news or disinformation, we assume the commonly used term ``misinformation’’ to encompass different aspects of misleading content. 

We acknowledge that imposing large monetary penalties on internet companies such as an information pollution tax requires a careful deliberation on what constitutes as information pollution, and that often is a democratic task that might have different outcomes depending on the participants. We intentionally leave the interpretation of ``misinformation'' to readers and operate on a ``know it when we see it'' basis in regards to what counts as information pollution. Future open discussions are needed to further clarify and define the boundaries between pollution and safe information.

\section{Taxing information pollution}
As previously noted in the carbon tax literature (e.g. \cite{metcalf2009design}), it is considered “heroic” to estimate the optimal tax rate (the marginal abatement cost and benefit) as defined by Pigou’s theory for CO2 and/or other greenhouse emissions, since calculating the optimal tax rate requires prediction about complex effects of climate change in the future. An alternative to calculating the optimal tax rate is to determine a set of taxes that overtime help to meet target emissions reduction goals and it has been shown that the two approaches converge to similar outcomes in practice \cite{metcalf2009design}. As misinformation is a more complex and less understood phenomenon compared to greenhouse emissions- and at times even fact-checkers struggle to identify fact from fiction, we also propose to set misinformation taxes by setting a misinformation reduction or information health target, rather than attempting to come up with accurate estimations of information pollution. 

To meet our information health target through taxation of misinformation, we need to understand two fundamental questions: (i) How much misinformation exists on each internet platform? and (ii) What are the economic, social, and human costs of misinformation? 

It requires far more research than just one paper to gain a deeper understanding of the two questions, so throughout the rest of the paper we aim to focus on raising the important questions and suggest high level directions that could eventually support operationalizing such a mechanism, if one is ever to be implemented.

To address the first question, we propose to focus on viral misinformation, misleading narratives that have reached more than some threshold of user engagement. Fringe misinformation is less likely to cause real damage and even fact-checking fringe misinformation might give such narratives more recognition they would normally receive \cite{phillips2018oxygen, wardle2019comprova}. Platforms already compute engagement metrics, and with the recently expanding fact-checking networks across the internet, keeping track of viral misinformation on each platform becomes a matter of record keeping that can be used to report aggregate engagement information (provided through the platforms.) We sketch out the important details regarding question (ii) in the following section.

\section{The stakeholders and implications of misinformation}
There are no good public estimates of the cost of misinformation on individuals, society, the economy, and public health. In this section we take a high level view of this complex ecosystem and demonstrate the extent to which misinformation affects humans. We aim for more breadth, and we identify measuring economic impacts of misinformation as an important future research and governance direction.

\subsection{The \textit{stakeholders}}

\begin{itemize}
    \item \textbf{Online media.}
    The owners of internet platforms stand to profit more from misinformation and controversy than from regular content. \cite{vosoughi2018spread} They also experience political backlash because of online misinformation, which in effect imposes public relations related costs on platforms. Since online media control the apparatus of information circulation, they are the most leveraged of all stakeholders.
    
    \item \textbf{Advertisers.}
    The vast majority of revenue of online media is through the sale of user attention to advertisers. This gives advertisers noticeable power over the platforms, and therefore they can exert force on internet companies using their spending as leverage. In the United States' traditional broadcasting and news media, the advertisers’ lobbying power has led the Federal Communications Commission (FCC)’s to regulate “obscenity, indecency, and profanity” \cite{federal2017obscene}, and fine media companies for publishing such content. Advertisers may choose to follow similar paths in response to online misinformation tainting their brands. According to Media Matters' report \cite{muskads}, Twitter lost half of its top 100 advertisers- who purchased nearly \$2 billion worth of ads since 2020, only a month after Elon Musk acquired the company. The advertisers included large corporations such as American Express, Citigroup, Chipotle, Nestle, Black Rock, and Chanel, and some publicly cited controversies and concerns around looser content moderation post Musk take over.
    
    \item \textbf{Journalists and fact-checkers.}
    At the forefront of reporting on world events, journalists and fact-checking communities often have to spend extra time and energy to go against misinformation, by avoiding them in reporting and doing extra work debunking falsehoods. 
    Journalism has become a more challenging and risky profession in recent years in part as a side effect of excessive information pollution.
    It is worth noting that the rivalry among traditional news media and online media for user attention is a confounding factor that sometimes interferes with journalism's impartiality in promoting the best course of action against misinformation. 
    Nevertheless, journalists and fact-checkers remain our best source of professional and expert advocacy against misinformation, since they have an economic interest in protecting their work from falsehoods.

    \item \textbf{Civil society.}
    Misinformation can have far-reaching and harmful effects on civil society. When people are exposed to false or misleading information, they may form incorrect opinions and beliefs that can harm the social fabric of a community. For example, misinformation can create divisions among people by fueling prejudice, mistrust, and hate. It can also erode public trust in institutions, such as the government, media, and scientific community, which are critical for maintaining a healthy democracy. Misinformation can also lead to the spread of harmful practices and ideologies. For example, false information about vaccine safety can discourage people from getting vaccinated, leading to the spread of preventable diseases. Similarly, misinformation about climate change can undermine efforts to address this pressing global issue. In addition, misinformation can also have serious consequences for individual and collective decision-making. People who rely on false information to make decisions may end up taking actions that are not in their best interest or the interest of society as a whole. Therefore civil society and grassroots organizations are an important lobby against misinformation.
\end{itemize}

While at first glance it seems that all parties must want to combat misinformation, they collectively have struggled to do so over the years. Social media owners prioritize short-term profits over the potential long-term risks of misinformation to their business. The architecture of online advertisement affords the advertisers to turn a blind eye and collect profits, since ads are targeted towards demographics and are not attached to content. The former two stakeholders possess the power, but lack the will to make structural changes to control misinformation. The other stakeholders, mainly civil society and journalists are in the reverse position, as they do not possess as much control over the flow of information, but have demonstrated interest and will to counter misinformation. As long as no governing entity addresses this power imbalance among stakeholders by applying appropriate regulation, we face the risks of widespread misinformation.

\subsection{Social cost of misinformation}
To understand the magnitude of misinformation’s social debt, we turn to prior research and reporting that associate a monetary value with negative outcomes of information pollution.

In a 2019 report \cite{cavazos2019economic}, a group of economists and cybersecurity analysts placed a \$78 billion price tag on the global damages of fake news, citing the most affected sectors as \textit{stock market losses and volatility} (\$39 billion), \textit{financial misinformation} (\$17 billion in US alone), and \textit{reputation management} and \textit{public health} (US only) costing an annual \$9 billion each, all in a single year. According to research from the Economic Policy Institute published in 2017, retirement savers lose an annual \$17 billion from acting on misleading advice from financial advisors with conflicts of interest. \cite{shierholz2017here, retirementsavings}

Additionally, activists and non-profit organizations have mobilized in recent years to study the finances driving online misinformation, and at times have successfully demonetized the interests behind pushing misleading narratives. Sleeping Giants are an activist organization comprised of mostly anonymous members with active chapters in the US, Australia, Brazil, Canada, France, and Germany. Since their inception in late 2016, they have successfully demonetized extremist and fake news websites as famous as Breitbart News, causing 820 corporations including AT\&T, BMW, and Visa to stop advertising with the far right outlet. \cite{boycottbreitbart}
Global Disinformation Index (GDI)\footnote{\url{https://www.disinformationindex.org}} is a not-for-profit organization that publishes open research on news markets around the world. GDI provides dynamic exclusion lists of global news organizations rated high risk for misinformation to adtech companies, effectively providing a mechanism for systematically defunding misinformation.

In late 2022 the families of the victims of the Sandy Hook elementary school massacre which occurred ten years prior, won two defamation cases against Alex Jones, the infamous conspiracy theorist that circulated baseless lies about the victims being hired actors by the government in a conspiracy to take away Americans' guns. Jones has been ordered to pay \$1.49 billion in damages in two Sandy Hook defamation cases, and awaits a third trial pending investigation in Texas. \cite{jonessandy} The ruling is a first of its kind in the United States, setting precedent in punishment for defamation through deploying misinformation.

Further research and inquiry is required to draw a clearer picture of the burdens information pollution imposes on society. We highlight a reoccurring theme in the reports we reference on economic damages of misinformation: the estimates surpass billions of USD, and the damages often directly affect the public. We believe these numbers are alarming enough to justify taxing internet platforms for information pollution.

\subsection{Public health implications of misinformation}
Misinformation can have serious public health implications, as it can spread false or misleading information about health issues, treatments, and interventions. Some of the key public health implications of misinformation include:

\begin{itemize}
    \item \textbf{Discouraging vaccination.} False information about vaccine safety can discourage people from getting vaccinated, leading to the spread of preventable diseases. This can have serious consequences for public health, especially in the context of outbreaks and pandemics. There is a growing body of research studying the impact of misinformation on vaccination and public health. \cite{loomba2021measuring, lee2022misinformation, mutombo2022covid, garett2021online}

    \item \textbf{Promoting risky treatments.} Misinformation about health treatments can lead people to seek out dangerous or ineffective remedies, which can be harmful to their health. For example, false information about the dangers of conventional medical treatments can lead people to rely on unproven alternative therapies. \cite{xu2021tiktok, loeb2020fake, loeb2022representation}

    \item \textbf{Undermining public trust in science.} Misinformation about health issues can erode public trust in science and scientific institutions.\cite{goldstein2020science, doi:10.1073/pnas.1912444117, doi:10.1073/pnas.1805871115} This can make it difficult for public health authorities to effectively communicate important health information and promote evidence-based practices. \cite{office2021confronting, doi:10.1056/NEJMp2207252}

    \item \textbf{Delaying treatment.} False information about symptoms and treatments can lead people to delay seeking medical help, which can have serious consequences for their health. For example, false information about the causes of cancer can discourage people from seeking early detection and treatment, which can reduce the chances of successful treatment.
    
\end{itemize}

Such implications even at small scales pose huge risks to community and public health as social changes such as anti-vaccination movements can cause exponentially worse public health outcomes, leading the US surgeon general to declare misinformation as a public health emergency. \cite{office2021confronting} Many parts of the healthcare industry including doctors, nurses, and medical staff are impacted by misinformation in a variety of ways such as:

\begin{itemize}
    \item \textbf{Healthcare resource allocation.} Misinformation can lead to an overuse or misuse of healthcare resources. For example, people may seek unnecessary medical treatments or tests based on false information, which can strain healthcare systems and divert resources away from those who need it most.

    \item \textbf{Health system costs.} The unnecessary overuse of healthcare resources caused by exposure to information pollution can increase healthcare costs, which can negatively impact patients, hospitals, and governments.

    \item \textbf{Burnout of hospital staff.} Hospital staff, including doctors, nurses, and support staff, may experience stress and burnout as they work to manage the consequences of misinformation. For example, they may have to spend extra time educating patients about accurate health information or addressing the fallout from misinformation-driven health decisions. \cite{doi:10.1056/NEJMp2207252}
    
\end{itemize}

While there are arguments both for and against attaching a monetary value to pain and suffering \cite{valueoflife, kling2012exxon} of patients and hospital staff, we find a “fine” designation to be appropriate for deterring the negative impact of health-related misinformation as a sound and promising regulatory path.

\section{Considerations for a misinformation tax}

The negative externalities of misinformation on social media platforms adds to a growing number of concerns regarding the lack of accountability from big tech corporations that profit from information pollution. Information pollution will likely exacerbate due to the recent influx in misleading AI-generated content on the internet \cite{NEURIPS2019_3e9f0fc9}. In a recent U.S. Supreme Court hearing on ``Gonzalez v. Google LLC\footnote{\url{https://www.supremecourt.gov/docket/docketfiles/html/public/21-1333.html}},'' the plaintiff argues that internet companies such as Google should no longer be exempt from liability under section 230 of the Communications Decency Act of 1996,\footnote{\url{https://www.fcc.gov/general/telecommunications-act-1996}} as personalized and targeted recommendations made by search engines and social media are \textit{intentionally} selective about what content is displayed to users. Our arguments in this paper align with that of the defense, and we believe that a Pigouvian tax on misinformation has the potential to address the issues of content moderation in the absence of such publisher protections for internet companies.

\subsection{Implementation considerations} 

How platforms respond to these taxes will be different: some may start charging toxic users, others might move away from the online advertising-attention market, some might ban information pollution while others ban pollutants. We find the variability in platforms' response as a positive outcome, since it leaves room for platforms to make decisions that makes sense for their businesses. Since “one size fits all” solutions might succeed in one place but fail in others, the flexibility afforded by misinformation tax to platforms in choosing how to address misinformation is noteworthy.

Online media manifests in different shapes and forms. One particularly challenging-to-oversee category of online media are closed messaging social media platforms such as WhatsApp, Signal, and Telegram. Due to end-to-end encryption, such platforms have no effective way to review content, as the content only exists on end-user devices in encrypted format. While WhatsApp misinformation may be less prevalent in Western countries, billions of users in India, Brazil, and Iran make heavy use of WhatsApp and Telegram. A promising recent response to this issue called ``tipline'' \cite{kazemi2022research} enables opt-in inquiry and reporting about potentially misleading claims users come across on the closed platforms. Tiplines have been found to uncover a significant portion of viral misinformation on WhatsApp public group chats before they went viral, in a case study conducted on the 2019 Indian general elections. \cite{kazemi2022research}

\subsection{Ethical considerations}
Freedom of expression is a fundamental human right and stopping misinformation should not turn into censorship. Platforms often cite similar concerns explaining controversial takedowns \cite{zuckfreedom}. A misinformation tax directed at social media platforms avoids regulating speech at the government level and affords flexible moderation mechanisms to platforms while disincentivizing misleading consequential online interactions at a macro level. From this perspective, misinformation tax is a light-handed regulatory measure that can be tailored to the specific needs of communities and governing bodies. On the downside, compliance with misinformation tax for platforms might mean making consequential changes to company procedures that reduce their edge against international competitors and affect small businesses on their platforms. However, an emergency discipline designation to information pollution \cite{doi:10.1073/pnas.2025764118} requires us to view these consequences as secondary priorities next to protecting individuals and the society against information pollution. Such regulation should be applied justly to all companies, and that might mean a larger tax base for foreign firms to account for issues of competitiveness. Furthermore, one might argue that if social media cannot survive without misinformation, this could be just cause to look for alternatives to current platforms as they are not economically productive at the macro level.

\section{Conclusion}
We drew analogies between misinformation and environmental pollution, and proposed similar regulatory frameworks--namely a Pigouvian tax on information pollution, to limit the spread of online misinformation.
In this paper, we present a high level plan for implementing ``misinformation tax.'' We acknowledge that this is only the tip of the iceberg and that adopting such proposals requires further research and considerations in practice, some of which will vary greatly depending on context. We identify macro-level estimations of both cost and reach of viral misinformation in online communities as an important direction for future work, while emphasizing the importance of soliciting community feedback and engagement in implementing such measures for reducing information pollution.

\begin{acks}
We thank Ashley Craig (Research School of Economics at Australian National University), Catherine Hausman (Ford School of Public Policy at University of Michigan), Eric Gilbert (School of Information at University of Michigan), and Isabelle Barroso (independent journalist) who provided feedback on earlier iterations of this work.
\end{acks}

\bibliographystyle{ACM-Reference-Format}
\bibliography{sample}

% \appendix

\end{document}